\documentclass[english,aps,reprint,superscriptaddress,groupedaddress]{revtex4}
\usepackage[T1]{fontenc}
\usepackage[utf8]{inputenc}
\usepackage{color}
\usepackage{babel}
\usepackage{dsfont}
\usepackage{amssymb}
\usepackage{graphicx}
\usepackage{caption}
\usepackage{subcaption}
\usepackage{esint}
\usepackage[bookmarks=false,breaklinks=false,pdfborder={0 0 1},backref=section,colorlinks=true]{hyperref}
\hypersetup{
 urlcolor=darkblue,citecolor=darkblue,linkcolor=darkred,hyperfootnotes=false}
\makeatletter

\@ifundefined{textcolor}{}
{%
 \definecolor{BLACK}{gray}{0}
 \definecolor{WHITE}{gray}{1}
 \definecolor{RED}{rgb}{1,0,0}
 \definecolor{GREEN}{rgb}{0,1,0}
 \definecolor{BLUE}{rgb}{0,0,1}
 \definecolor{CYAN}{cmyk}{1,0,0,0}
 \definecolor{MAGENTA}{cmyk}{0,1,0,0}
 \definecolor{YELLOW}{cmyk}{0,0,1,0}
}

\definecolor{darkblue}{rgb}{0,0,0.6}\definecolor{darkred}{rgb}{0.6,0,0}\usepackage{babel}
\makeatother

\begin{document}
\title{Schr\"odinger Bridges for Systems of Interacting Particles}
\date{\today}
\author{Henri Orland}
\date{\today}
\address{Institut de Physique Théorique ~\\
 CNRS, CEA ~\\
 Université Paris-Saclay ~\\
 France }
\begin{abstract}
Abstract: A Schr\"odinger bridge is the most probable time-dependent
probability distribution that connects an initial probability distribution
$w_{i}$ to a final one $w_{f}$. The problem has been solved and
widely used for the case of simple Brownian evolution (non-interacting
particles). It is closely related to the problem of entropy-regularized Wasserstein optimal transport.
In this article, we generalize the problem to systems of interacting
particles.
We derive some equations for the forward and backward single
particle ``wave-functions'', which allows computation of the most probable
evolution of the single-particle probability between the initial and
final distributions. 
\end{abstract}
\maketitle
\section*{Introduction}
In a seminal article published in 1931 \cite{Schrod1,Schrod2,Chetrite},
Schr\"odinger, noticing the analogy between his equation for the dynamics
of quantum particles and the diffusion equation, which describes the
statistics of independent particles performing Brownian motion, proposed
and solved the following problem. Consider two sets of probability
distributions $w_{i}(r)$ and $w_{f}(r)$. There is a finite probability
that an ensemble of particles with distribution $w_{i}(r)$ at time
0, evolves under a Brownian motion to the distribution $w_{f}(r)$
at time $t_{f}$. How can one determine the probability distribution
for the particles $w(r,t)$ at intermediate times $t$ ? This is called
the \emph{Schr\"odinger bridge problem} and Schr\"odinger found an elegant {\em exact}
solution to this problem. It is a question of paramount importance
in non-equilibrium statistical physics, as it determines how an ensemble
of particles in a given density configuration evolves into another
one. For an extensive mathematical review, see \cite{Leonard}.

The Schr\"odinger bridge problem has been shown to be the dynamic version of 
the entropy regularized optimal transport problem with Wasserstein
distance \cite{Foellmer,OT}. 
Indeed, it was shown that the dynamical Schr\"odinger problem is equivalent to an optimal control problem where the cost function is the Kullback-Leibler divergence \cite{KL} between the given probability measures and a reference one (which is the free diffusion one).
 In addition, the optimal mass
transport exactly corresponds to the optimization of the entropy
production functional by an overdamped Langevin dynamics \cite{Aurell}.

 Due to this fact, the Schr\"odinger bridge problem
has found numerous applications in recent years in e.g. statistical
physics \cite{SP,Aurell,Saito}, transition path theory \cite{TP}, optimal transport
\cite{OT}, machine learning \cite{ML}, computer vision \cite{CV},
finances \cite{finan} etc. to name a few. The original problem
deals with independent free (Brownian) particles. The goal of this
article is to generalize the concept to the case of interacting particles.
Contrary to the free particle case, the interacting particle Sch\"odinger bridge does not have an explicit exact solution, since the Green's function for this problem cannot be computed explicitly. The purpose of this article is to introduce an exact field theoretical representation of the problem and to use techniques of field theory, such as the saddle-point method, to solve it. As we shall see, this field theory is naturally derived from the use of second-quantized coherent states.

\section*{The Model}
Consider a system of $N$ particles, interacting with a one-body potential
$U(r)$ and a two-body potential $v(r-r')$ . The initial coordinates
of the particles at $t=0$ are $\left\{ r_{1}^{i},...,r_{N}^{i}\right\} $
and their final coordinates at time $t_{f}$ are $\left\{ r_{1}^{f},...,r_{N}^{f}\right\} $.

These configurations define the probability distributions $w_{i}(r)$
and $w_{f}(r)$ by 
\begin{eqnarray}
w_{i}(r) & = & \frac{1}{N}\sum_{k=1}^{N}\delta(r-r_{k}^{i})\\
w_{f}(r) & = & \frac{1}{N}\sum_{k=1}^{N}\delta(r-r_{k}^{f})
\end{eqnarray}
and the corresponding particle densities by 
\begin{eqnarray}
\label{density}
\rho_{i}(r) & = & Nw_{i}(r)=\sum_{k=1}^{N}\delta(r-r_{k}^{i}) \nonumber\\
\rho_{f}(r) & = & Nw_{f}(r)=\sum_{k=1}^{N}\delta(r-r_{k}^{f})
\end{eqnarray}
In the following, we will often use the notation $\left\{ r_{i}\right\} $ to represent the set of vectors $\left( r_{1},...,r_{N}\right) $.
The particles evolve under overdamped Langevin dynamics \cite{LD,HO}
(Brownian dynamics)

\begin{equation}
\dot{r}_{i}=-\frac{1}{\gamma}\nabla_{i}W(\{r_{i}\})+\eta_{i}(t)
\end{equation}
where

\begin{equation}
W(\{r_{i}\})=\sum_{i}U(r_{i})+\frac{1}{2}\sum_{i\neq j}v(r_{i}-r_{j})
\end{equation}
and 
\begin{equation}
\nabla_{i}W(\{r_{i}\})=\nabla_{i}U(r_{i})+\sum_{j\neq i}\nabla_{i}v(r_{i}-r_{j})
\end{equation}
The friction coefficient $\gamma$ is related to the diffusion coefficient
$D$ and to the inverse temperature $\beta=1/k_{B}T$, by the Einstein
relation $\gamma=1/\beta D$ and $\eta_{i}(t)$ is a random Gaussian
force, with zero mean and correlation given by

\begin{equation}
\left\langle \eta_{i}(t)\eta_{j}(t')\right\rangle =2D\delta(t-t')
\end{equation}

The probability distribution for the particles $P(\left\{ r_{i}\right\} ,t)$
satisfies the Fokker-Planck equation \cite{LD,HO} 
\begin{equation}
\frac{dP(\left\{ r_{i}\right\} ,t)}{dt}=D\sum_{i}\nabla_{i}\left(\nabla_{i}P+\beta P\nabla_{i}W\right)
\end{equation}

Note that the duration $t_{f}$ of the process can be eliminated by
rescaling the time $t$ in the Langevin or Fokker-Planck equation,
provided the diffusion constant and the temperature are rescaled properly.
Most of the mathematical literature uses $t_{f}=1$. Since, in this
article, we emphasize the physical aspect of evolving particles, we
shall keep the original time $t_{f}$.

Defining the function

\begin{equation}
\Phi(\{r_{i}\},t)=e^{+\frac{\beta}{2}W(\{r_{i}\})}P(\{r_{i}\},t)
\end{equation}
it satisfies the Schr\"odinger equation \cite{Zwan,HO}

\begin{eqnarray}
\frac{d\Phi(\{r_{i}\},t)}{dt} & = & -H\Phi(\{r_{i}\},t)\\
 & = & D\sum_{i}\nabla_{i}^{2}\Phi-D\beta^{2}V_{eff}(r_{i})\Phi(\{r_{i}\},t)
\end{eqnarray}
where the Hamiltonian is given by

\begin{equation}
H=-D\sum_{i}\nabla_{i}^{2}+D\beta^{2}V_{eff}(\{r_{i}\})
\end{equation}
and the effective potential is given by

\begin{equation}
V_{eff}(\{r_{i}\})=\frac{1}{4}\sum_{i}(\nabla_{i}W)^{2}-\frac{k_{B}T}{2}\sum_{i}\nabla_{i}^{2}W
\end{equation}
which in terms of the potentials $U$ and $v$ takes the form

\begin{equation}
V_{eff}(\{r_{i}\})=\frac{1}{4}\sum_{i}\left(\nabla_{i}U(r_{i})+\sum_{j\neq i}\nabla_{i}v(r_{i}-r_{j})\right)^{2}-\frac{k_{B}T}{2}\sum_{i}\left(\nabla_{i}^{2}U(r_{i})+\sum_{j\neq i}\nabla_{i}^{2}v(r_{i}-r_{j})\right)
\end{equation}
Note that the first term is a 3-body interaction, whereas the second
is a 2-body one. %For the sake of notational simplicity, we will
%assume that there is no one-body potential, so that
%
%\begin{equation}
%V_{eff}(r_{i})=\frac{1}{4}\sum_{i}\sum_{j\neq i}\sum_{k\neq i}\nabla_{i}v(r_{i}-r_{j})\nabla_{k}v(r_{i}-r_{k})-\frac{k_{B}T}{2}\sum_{i,j}\nabla_{i}^{2}v(r_{i}-r_{j})
%\end{equation}
%

The probability that the set of particles starting at $\{r_{1}^{i},\ldots,r_{N}^{i}\}$
at time 0 ends in any permutation of the set $\{r_{1}^{f},\ldots,r_{N}^{f}\}$
at time $t_{f}$ is given by

\[
P(\{r_{k}^{f}\},t_{f}|\{r_{k}^{i}\},0)=\sum_{p\in S_{N}}P(r_{p(1)}^{f},...,r_{p(N)}^{f},t_{f}|r_{1}^{i},...,r_{n}^{i},0)
\]
where the summation is over all permutations $p$ in the symmetric
group $S_{N}$.

Using the quantum mechanical Dirac notation \cite{FH,QM,HO}, the
probability for the system to evolve from the initial configuration
to the final one is given by 
\begin{equation}
P(\{r_{k}^{f}\},t_{f}|\{r_{k}^{i}\},0)=\frac{1}{N!}e^{-\frac{\beta}{2}\left(W(\{r_{k}^{f}\})-W(\{r_{k}^{i}\})\right)}\left\langle r_{1}^{f},...,r_{N}^{f}\left|e^{-t_{f}H}\right|r_{1}^{i},...,r_{n}^{i}\right\rangle 
\end{equation}
where the brackets $|r_{1}\ldots r_{N}\rangle=\sum_{p\in S_{N}}|r_{p(1)}\ldots r_{p(N)})$
denote the symmetrized (bosonic) state summed over all permutations
$p$ of the symmetric group $S_{N}$ . The factor $1/N!$ cancels
one of the summations over permutations since the matrix element above
comprises two such summations. In terms of the particle density distributions
$\rho_{i}$ and $\rho_{f}$ , the pre-factor above can be written
as

\begin{eqnarray}
W(\{r_{k}^{f}\})-W(\{r_{k}^{i}\}) & = & \int drU(r)(\rho_{f}(r)-\rho_{i}(r))\nonumber \\
 & + & \frac{1}{2}\int drdr'v(r-r')(\rho_{f}(r)\rho_{f}(r')-\rho_{i}(r)\rho_{i}(r'))
\end{eqnarray}
where the $\rho$'s are related to the $\{ r_k\}$ through the relations (\ref{density}).
\section*{Functional Integral Representation}
In the following of this paper, we will heavily rely on coherent states
\cite{Neg_HO}. We recall that coherent states constitute an overcomplete
set of vectors in the Fock space of the system. The two fundamental
identities that allow calculations with coherent states $|\phi\left\rangle \right.$ \cite{Neg_HO} are \hfill
\break i)the scalar product 
\begin{eqnarray}
\left\langle \phi\left|\right.\phi'\right\rangle =e^{\int dr\phi^{*}(r)\phi'(r)}
\end{eqnarray}
\hfill \break
ii) the decomposition of unity 
\begin{eqnarray}
\label{closure}
\int d\phi^{*}(r)d\phi(r)e^{-\int dr\phi^{*}(r)\phi(r)}\left|\phi\right\rangle \left\langle \phi\right| & = & \mathds{1}
\end{eqnarray}
where the notation $\int d\phi(r)d\phi^{*}(r)$ denotes a functional integral over
the complex conjugate fields $\phi(r)$ and $\phi^{*}(r)$  and
$\mathds{1}$ is the unit operator defined by $\left\langle r\left|\mathds{1}\right|r'\right\rangle =\delta(r-r')$.
We remind the reader that the functional integral over the complex fields $\phi(r)$ and $\phi^{*}(r)$ can be understood in the following way: space is discretized on a lattice with $M$ points and lattice spacing $a$, and the fields are complex integration variables defined at each point of the lattice. 
One then takes the continuous limit of space $M \to \infty, \ a \to 0$, which leads to an infinite number of complex integrals. As a result, infinities occur, and it is the purpose of {\em renormalization theory} to tackle these infinities \cite{Peskin}. However, at the level of the saddle-point approximation, which we will use throughout this article, such infinities do not occur.
\break
{So for instance equation (\ref{closure}) is to be undestood as
\begin{equation}
\lim_{M \to \infty, a \to 0} \int \prod_{i=1}^M \frac{d\phi^*_i d\phi_i}{2 \pi} e^{- a^d \sum_{i=1}^M \phi^*_i \phi_i} | \phi_1...\phi_M\rangle  \langle \phi_1...\phi_M| = \mathds{1}
\end{equation}
where $d$ is the spatial dimension of the system.

\noindent The notation
\begin{equation}
d\phi^*(r) d\phi(r) = \lim_{M \to \infty , a\to 0} \prod_{i=1}^M \frac{d\phi^*_i  d \phi_i}{2 \pi}
\end{equation}
will be used throughout the rest of this article.
}

\noindent
In addition, we have
\begin{eqnarray}
\left\langle r_{1}...r_{N}\right|\phi\left.\right\rangle  & = & \phi(r_{1})...\phi(r_{N})\\
\left\langle \phi\left|r_{1}...r_{n}\right.\right\rangle  & = & \phi^{*}(r_{1})...\phi^{*}(r_{N})
\end{eqnarray}
Using these identities, we have

\begin{eqnarray}
\left\langle r_{1}^{f},...,r_{N}^{f}\left|e^{-t_{f}H}\right|r_{1}^{i},...,r_{n}^{i}\right\rangle  & = & \int d\phi_{f}^{*}(r)d\phi_{f}(r)\int d\phi_{i}^{*}(r)d\phi_{i}(r)e^{-\int dr\left(\phi_{f}^{*}(r)\phi_{f}(r)+\phi_{i}^{*}(r)\phi_{i}(r)\right)}\nonumber \\
 & \times & \phi_{f}(r_{1}^{f})...\phi_{f}(r_{N}^{f})\phi_{i}^{*}(r_{1}^{i})...\phi_{i}^{*}(r_{N}^{i})\left\langle \phi_{f}\left|e^{-t_{f}H}\right|\phi_{i}\right\rangle 
\end{eqnarray}
The coherent state matrix element above can be represented as a functional
integral over complex fields \cite{Neg_HO} as 
\begin{eqnarray}
\left\langle \phi_{f}\left|e^{-t_{f}H}\right|\phi_{i}\right\rangle  & = & \int_{\phi(r,0)=\phi_{i}(r)}^{\phi^{*}(r,t_{f})=\phi_{f}^{*}(r)}\mathcal{D}(\phi^{*},\phi)\exp\Bigg\{\int dr\phi_{f}^{*}(r)\phi_{f}(r)\nonumber \\
 & - & \int_{0}^{t_{f}}dt\int dr\left(\phi^{*}(r,t)\left(\frac{\partial\phi}{\partial t}-D\nabla^{2}\right)\phi(r,t)\right)-\int_{0}^{t_{f}}dtA(\phi^{*},\phi)\Bigg\}\nonumber \\
\end{eqnarray}
where the interaction term is given by

\begin{eqnarray}
A(\phi^{*},\phi) & = & D\beta^{2}\int dr\left|\phi(r,t)\right|^{2}\Bigg(\frac{1}{4}\left(\nabla U(r)+\int dr'\nabla v(r-r')\left|\phi(r',t)\right|^{2}\right)^{2}\label{inter}\nonumber \\
 & - & \frac{k_{B}T}{2}\Big(\nabla^{2}U(r)+\int dr'\nabla^{2}v(r-r')\left|\phi(r',t)\right|^{2}\Big)\Bigg)\\
 & = & D\beta^{2}\int dr\rho(r,t)\Bigg(\frac{1}{4}\left(\nabla U(r)+\int dr'\nabla v(r-r')\rho(r',t)\right)^{2}\\
 & - & \frac{k_{B}T}{2}\Big(\nabla^{2}U(r)+\int dr'\nabla^{2}v(r-r')\rho(r',t)\Big)\Bigg)
\end{eqnarray}
where

\[
\rho(r,t)=\phi^{*}(r,t)\phi(r,t)=\left|\phi(r,t)\right|^{2}
\]
This term can be rewritten as

\begin{equation}
A(\phi^{*},\phi)=D\int dr\rho(r,t)e^{\frac{\beta}{2}\left(U(r)+\int dr'v(r-r')\rho(r',t)\right)}\nabla^{2}e^{-\frac{\beta}{2}\left(U(r)+\int dr'v(r-r')\rho(r',t)\right)}\label{eq:A}
\end{equation}
Using the identity

\begin{eqnarray}
\phi_{f}(r_{1}^{f})...\phi_{f}(r_{N}^{f})\phi_{i}^{*}(r_{1}^{i})...\phi_{i}^{*}(r_{N}^{i}) & = & \exp\left(\sum_{k}\left(\log\phi_{f}(r_{k}^{f})+\log\phi_{i}^{*}(r_{k}^{i})\right)\right)\nonumber \\
 & = & \exp\left(\int dr\left(\rho_{f}(r)\log\phi_{f}(r)+\rho_{i}(r)\log\phi_{i}^{*}(r)\right)\right)\nonumber \\
\end{eqnarray}
we finally obtain the expression for the probability for the system
to make a transition from the probability distribution $w_{i}(r)$
to $w_{f}(r)$ during time $t_{f}$ :

\begin{eqnarray}
P(w_{f},t_{f}\left|w_{i},0\right.) & = & \frac{1}{N!}\exp\Big(-\frac{\beta}{2}\int drU(r)(\rho_{f}(r)-\rho_{i}(r))\nonumber \\
 & - & \frac{\beta}{4}\int drdr'v(r-r')(\rho_{f}(r)\rho_{f}(r')-\rho_{i}(r)\rho_{i}(r'))\Big)\nonumber \\
 & \times & \int\mathcal{D}(\phi^{*},\phi)\exp\Bigg\{-\int dr\phi^{*}(r,0)\phi(r,0)\nonumber \\
 & - & \int_{0}^{t_{f}}dt\int dr\left(\phi^{*}(r,t)\left(\frac{\partial\phi}{\partial t}-D\nabla^{2}\right)\phi(r,t)\right)-\int_{0}^{t_{f}}dtA(\phi^{*},\phi)\nonumber \\
 & + & \int dr\left(\rho_{f}(r)\log\phi(r,t_{f})+\rho_{i}(r)\log\phi^{*}(r,0)\right)\Bigg\}\label{proba}
\end{eqnarray}
where $A(\phi^{*},\phi)$ is given by (\ref{inter}).

Note that the boundary conditions on the functional integral are not
present anymore, since $\phi_{f}^{*},\phi_{i}$ are fully integrated
over. To complement these equations, we note that the conditioned
density $\rho_{c}(r,t)$ of particles at point $r$ at time $t$ is
given by

\begin{equation}
\rho_{c}(r,t)=\left\langle \left|\phi(r,t)\right|^{2}\right\rangle \label{cond}
\end{equation}
where the bracket denotes the expectation value with the probability
defined in eq. (\ref{proba})

\subsection*{The one-body (non-interacting) case}

We first consider the free case $v=0$. In that case, there is no
interaction term and the probability becomes

\begin{eqnarray}
P(w_{f},t_{f}\left|w_{i},0\right.) & = & \frac{1}{N!}\exp\left(-\frac{\beta}{2}\int drU(r)(\rho_{f}(r)-\rho_{i}(r))\right)\nonumber \\
 & \times & \int\mathcal{D}(\phi^{*},\phi)\exp\Bigg\{-\int dr\phi^{*}(r,0)\phi(r,0)\\
 & - & \int_{0}^{t_{f}}dt\int dr\left(\phi^{*}(r,t)\left(\frac{\partial\phi}{\partial t}-D\nabla^{2}+\frac{D\beta^{2}}{4}\left(\left(\nabla U(r)\right)^{2}-2k_{B}T\nabla^{2}U(r)\right)\right)\phi(r,t)\right)\nonumber \\
 & + & \int_{0}^{t_{f}}dr\left(\rho_{f}(r)\log\phi(r,t_{f})+\rho_{i}(r)\log\phi^{*}(r,0)\right)\Bigg\}
\end{eqnarray}

\noindent Since this functional integral is Gaussian, the saddle-point
approximation is exact. Note that although the fields $\phi$ and
$\phi^{*}$ are complex conjugate in the functional integral, in the
saddle-point approximation, they no longer need to be complex conjugate
from each other, and they are to be treated as two independent complex
fields.

The equations extremizing the exponent read:

\begin{eqnarray}
\left(\frac{\partial}{\partial t}-D\nabla^{2}+\frac{D\beta^{2}}{4}\left(\left(\nabla U(r)\right)^{2}-2k_{B}T\nabla^{2}U(r)\right)\right)\phi(r,t) & = & 0\label{SPE1}\\
\left(-\frac{\partial}{\partial t}-D\nabla^{2}+\frac{D\beta^{2}}{4}\left(\left(\nabla U(r)\right)^{2}-2k_{B}T\nabla^{2}U(r)\right)\right)\phi^{*}(r,t) & = & 0\label{SPE2}
\end{eqnarray}
with the boundary conditions (obtained by extremizing the exponent
w.r.t. the fields at 0 and $t_{f})$ 
\begin{eqnarray}
\phi(r,0)=\frac{\rho_{i}(r)}{\phi^{*}(r,0)}\\
\phi^{*}(r,t_{f})=\frac{\rho_{f}(r)}{\phi(r,t_{f})}\\
\end{eqnarray}
Note that (\ref{SPE2}) is the time-reversed equation of (\ref{SPE1}).
However, the boundary conditions above prevent the expression of $\phi^{*}(r,t)$
in terms of $\phi(r,t)$.

{We define the Green's function $G(rt,r't')$ as the inverse of the operator
\begin{eqnarray}
\frac{\partial}{\partial t}-D\nabla^{2}+\frac{D\beta^{2}}{4}\left(\left(\nabla U(r)\right)^{2}-2k_{B}T\nabla^{2}U(r)\right)
\end{eqnarray}
It is a solution of the equation 
}
\begin{equation}
\left(\frac{\partial}{\partial t}-D\nabla^{2}+\frac{D\beta^{2}}{4}\left(\left(\nabla U(r)\right)^{2}-2k_{B}T\nabla^{2}U(r)\right)\right)G(rt,r't')=\delta(r-r')\delta(t-t')
\end{equation}
In the absence of an external potential $U$, in space dimension $d$,
this Green's function reads 
\begin{equation}
G_{0}(rt,r't')=\left(\frac{1}{4\pi D(t-t')}\right)^{d/2}e^{-\frac{(r-r')^{2}}{4D(t-t')}}
\end{equation}
Another simple case when this Green's function can be computed exactly
is that of the harmonic oscillator (Ornstein-Uhlenbeck process) 
\begin{equation}
U(r)=\frac{1}{2}Kr^{2}
\end{equation}
%since the effective potential is also quadratic 
%\begin{equation}
%(\nabla U(r))^{2}-2k_{B}T\nabla^{2}U=K^{2}r^{2}-2k_{B}TK
%\end{equation}

We introduce the notation

\begin{eqnarray}
\phi_{i}(r) & = & \phi(r,0)=\frac{\rho_{i}(r)}{\phi^{*}(r,0)}\label{eq:BC}\\
\phi_{f}^{*}(r) & = & \phi^{*}(r,t_{f})=\frac{\rho_{f}(r)}{\phi(r,t_{f})}
\end{eqnarray}
Eq. (\ref{SPE1},\ref{SPE2}) with their boundary conditions can be
solved as

\begin{eqnarray*}
\phi(r,t) & = & \int dr'G(rt,r'0)\phi_{i}(r')\\
\phi^{*}(r,t) & = & \int dr'\phi_{f}^{*}(r')G(r't_{f},rt)
\end{eqnarray*}
which implies 
\begin{eqnarray}
\rho_{f}(r) & = & \phi_{f}^{*}(r)\int dr'G(rt_{f},r'0)\phi_{i}(r')\label{SC1}\\
\rho_{i}(r) & = & \phi_{i}(r)\int dr'\phi_{f}^{*}(r')G(r't_{f},r0)\label{SC2}
\end{eqnarray}
These are the standard Schr\"odinger bridge equations. They are the
same equations as those of the entropy-regularized optimal transport
problem \cite{OT}, with the Wassestein cost function $(r-r')^{2}/4t_{f}$.
Therefore, they can be solved efficiently by using the Sinkhorn fixed-point
method \cite{Sink}. Starting from a guess for $\phi_{i}(r)$, we
obtain $\phi_{f}^{*}(r)$ from eq.(\ref{SC1}). We plug it in (\ref{SC2})
to obtain a new $\phi_{i}(r)$, and the procedure is iterated until
convergence. Convergence is guaranteed and very fast (\cite{Conv}).

From eq.(\ref{cond}) and since in the saddle-point approximation
the expectation value is identical to the actual value, the conditioned
particle density is given by 
\begin{eqnarray*}
\rho_{c}(r,t) & = & \left|\phi(r,t)\right|^{2}\\
 & = & \int dr'dr''\phi_{f}^{*}(r')G(r't_{f},rt)G(rt,r''0)\phi_{i}(r'')
\end{eqnarray*}
which satisfies the boundary conditions $\rho_{c}(r,0)=\rho_{i}(r)$
and $\rho_{c}(r,t_{f})=\rho_{f}(r)$.

Finally, the transition probability is given by 
\begin{eqnarray}
P(w_{f},t_{f}\left|w_{i},0\right.) & = & N!\exp\left(-\frac{\beta}{2}\int drU(r)(\rho_{f}(r)-\rho_{i}(r))\right)\nonumber \\
 & \times & \exp\left(N\int dr\left[w_{i}(r)\log\frac{w_{i}(r)}{\phi_{i}(r)}+w_{f}(r)\log\frac{w_{f}(r)}{\phi_{f}^{*}(r)}\right]\right)\nonumber \\
\end{eqnarray}
%which can be written as 
%\begin{eqnarray}
%P(w_{f},t_{f}\left|w_{i},0\right.) & = & N!^2\exp\left(-\beta\int drU(r)(\rho_{f}(r)-\rho_{i}(r))\right)\nonumber \\
% & \times & \exp\left(\int dr\left[\rho_{i}(r)\log\frac{\rho_{i}(r)}{\phi_{i}(r)}+\rho_{f}(r)\log\frac{\rho_{f}(r)}{\phi_{f}^{*}(r)}\right]\right)\nonumber \\
%\end{eqnarray}
We recognize the Kullback-Leibler (KL) relative entropy \cite{KL}
of the distributions $\phi_{i}$ and $\phi_{f}^{*}$ with respect
to the distributions $w_{i}$ and $w_{f}$. This expression is an illustration of 
the connection of Schr\"odinger bridges with large deviations theory
and optimal transport.

%We illustrate the free case with an exactly solvable example, namely
%that of an ``echo''. In that case, the final density is equal to
%the original one, $\rho_{f}(r)=\rho_{i}(r)$. We consider the one-dimensional
%case when the initial particles occupy the right half space $z>0$
%at time $0$, and go back to the same density at time $t_{f}$. Namely
%
%\[
%\rho_{f}(z)=\rho_{i}(z)=\theta(z)
%\]
% Due to the boundary conditions (\ref{eq:BC}), we assume the following
%form for $\phi_{i}=\phi_{f}^{*}$
%
%\begin{eqnarray*}
%\phi_{i}(z) & = & f(z)\theta(z)
%\end{eqnarray*}
%
%

\subsection*{The interacting case}

We now consider the interacting case. For the sake of simplicity,
we shall assume that $U=0$. The expression for $P$ is still given
by eq.(\ref{proba}), however the expression (\ref{inter}) simplifies
as 
\begin{eqnarray}
A(\phi^{*},\phi) & = & D\beta^{2}\int dr\rho(r,t)\left(\frac{1}{4}\left(\int dr'\nabla v(r-r')\rho(r',t)\right)^{2}-\frac{k_{B}T}{2}\int dr'\nabla^{2}v(r-r')\rho(r',t)\right)\nonumber \\
 & = & D\int dr\rho(r,t)e^{\frac{\beta}{2}\int dr'v(r-r')\rho(r',t)}\nabla^{2}e^{-\frac{\beta}{2}\int dr'v(r-r')\rho(r',t)}
\end{eqnarray}
where $\rho(r,t)=\left|\phi(r,t)\right|^{2}=Nw(r,t)$, and $w(r,t)$
is the probability distribution at time $t$.

Due to the presence of implicit factors of $N$ in $A$ through $\rho(r,t)$,
the saddle-point approximation is not exact. However, it provides
{\em the most probable probability distribution path} connecting $w_{i}$
and $w_{f}$,  which is specifically the problem posed by Schr\"odinger \cite{Schrod1}.

The saddle point equations are 
\begin{eqnarray}
 & \Bigg( & \frac{\partial}{\partial t}-D\nabla^{2}-D\beta\int dr'\nabla^{2}v(r-r')\rho(r',t)+\frac{D\beta^{2}}{4}\left(\int dr'\nabla v(r-r')\rho(r't)\right)^{2}\nonumber \\
 & + & \frac{D\beta^{2}}{2}\int dr'dr''\rho(r',t)\nabla v(r'-r'')\rho(r'',t)\nabla v(r'-r)\Bigg)\phi(r,t)=0\label{eq:sp1}\\
 & \Bigg( & -\frac{\partial}{\partial t}-D\nabla^{2}-D\beta\int dr'\nabla^{2}v(r-r')\rho(r',t)+\frac{D\beta^{2}}{4}\left(\int dr'\nabla v(r-r')\rho(r',t)\right)^{2}\nonumber \\
 & + & \frac{D\beta^{2}}{2}\int dr'dr''\rho(r',t)\nabla v(r'-r'')\rho(r'',t)\nabla v(r'-r)\Bigg)\phi^{*}(r,t)=0\label{eq:sp2}
\end{eqnarray}
with the boundary conditions 
\begin{eqnarray}
\phi(r,0)=\frac{\rho_{i}(r)}{\phi^{*}(r,0)}\label{int_BC1}\\
\phi^{*}(r,t_{f})=\frac{\rho_{f}(r)}{\phi(r,t_{f})}\label{int_BC2}
\end{eqnarray}
As mentioned in the non-interacting case, (\ref{eq:sp2}) is the time-reversed
equation of (\ref{eq:sp1}).

As before, the conditioned density of particles at $r$ at time $t$
is given by 
\[
\rho_{c}(r,t)=\left|\phi(r,t)\right|^{2}
\]
As in the non-interacting case, these equations can be solved numerically
by a Sinkhorn-like fixed-point method. One first guesses the function
$\rho(r,t)$ , with the condition $\rho(r,0)=\rho_{i}(r)$ and $\rho(r,t_{f})=\rho_{f}(r)$
. For example, one can use $\rho(r,t)=\frac{t_{f}-t}{t_{f}}\rho_{i}(r)+\frac{t}{t_{f}}\rho_{f}(r)$
as an initial guess. One must also guess $\phi(r,0)$. With this two
functions, one can solve (\ref{eq:sp1}) to obtain $\phi(r,t_{f})$.Using
(\ref{int_BC2}), we obtain $\phi^{*}(r,t_{f})$ which is used as
an initial condition for (\ref{eq:sp2}) to obtain $\phi^{*}(r,0)$.
Using (\ref{int_BC1}), we get $\phi(r,0)$ and we can recalculate
the density $\rho$ and iterate the procedure until convergence.

%Although these equations conserve the total number of particles (space
%integral of $\rho(r,t)$), it is crucial to normalize $\rho(r,t)$
%by requiring that its integral should be equal to that of $\rho_{i}$.
The above equations can be used to express the probability at the
saddle-point as

\begin{eqnarray*}
P(w_{f},t_{f}\left|w_{i},0\right.) & = & N!\exp\Big(-\frac{\beta}{4}\int drdr'v(r-r')(\rho_{f}(r)\rho_{f}(r')-\rho_{i}(r)\rho_{i}(r'))\Big)\\
\\ & \times & \exp\Bigg\{-\frac{D\beta}{2}\int_{0}^{t_{f}}dt\int drdr'\rho(r,t)\nabla^{2}v(r-r')\rho(r',t)\\
\\ & + & \frac{D\beta^{2}}{2}\int_{0}^{t_{f}}dt\int dr\rho(r,t)\left(\int\nabla v(r-r')\rho(r',t)\right)^{2}\\
 & + & N\int dr\left(w_{f}(r)\log\frac{w_{f}(r)}{\phi(r,t_{f})}+w_{i}(r)\log\frac{w_{i}(r)}{\phi^{*}(r,0)}\right)\Bigg\}
\end{eqnarray*}
Again, the KL relative entropy with respect to the original densities
appears in this expression.

\subsection*{One-dimensional example}

We now look at a specific one-dimensional example, namely 
\[
v(x)=g\delta(x)
\]
A positive $g$ corresponds to a repulsive interaction between the particles, whereas a negative one corresponds to attraction.
The interaction term (\ref{inter}) becomes 
\begin{eqnarray}
A(\phi^{*},\phi) & = & D\beta^{2}\int dx\rho(x,t)\Bigg(\frac{g^{2}}{4}\rho'{}^{2}(x,t)-\frac{g}{2}k_{B}T\rho''(x,t)\Bigg)%
\label{inter1}
\end{eqnarray}

Equations (\ref{eq:sp1}), (\ref{eq:sp2}) become 
\begin{eqnarray}
 & \Bigg( & \frac{\partial}{\partial t}-D\frac{\partial^{2}}{\partial x^{2}}-D\beta g\rho''(x,t)-\frac{D\beta^{2}g^{2}}{4}\rho'{}^{2}(x,t)\label{eq:1d1}\nonumber \\
 & - & \frac{D\beta^{2}g^{2}}{2}\rho(x,t)\rho''(x,t)\Bigg)\phi(x,t)=0\\
 & \Bigg( & -\frac{\partial}{\partial t}-D\frac{\partial^{2}}{\partial x^{2}}-D\beta g\rho''(x,t)-\frac{D\beta^{2}g^{2}}{4}\rho'{}^{2}(x,t)\label{eq:1d2}\nonumber \\
 & - & \frac{D\beta^{2}g^{2}}{2}\rho(x,t)\rho''(x,t)\Bigg)\phi^{*}(x,t)=0
\end{eqnarray}
with the boundary conditions

\begin{eqnarray}
\phi(x,0)=\frac{\rho_{i}(x)}{\phi^{*}(x,0)}\label{eq:1d3}\\
\phi^{*}(x,t_{f})=\frac{\rho_{f}(x)}{\phi(x,t_{f})}\label{eq:1d4}
\end{eqnarray}
%These equations can be solved numerically by a fixed-point method.
%One first guesses the function $\rho(x,t)$ , with the condition $\rho(x,0)=\rho_{i}(x)$
%and $\rho(x,t_{f})=\rho_{f}(x)$ . For example, one can use $\rho(x,t)=\frac{t_{f}-t}{t_{f}}\rho_{i}(x)+\frac{t}{t_{f}}\rho_{f}(x)$
%as an initial guess. One can also guess $\phi_{i}$ and $\phi_{f}$
%, for instance $\phi_{i}(x)=\rho_{i}(x)$ and $\phi_{f}^{*}(x)=\rho_{f}(x)$.
%One can then solve eq.(\ref{eq:1d1}) and (\ref{eq:1d2}) to obtain
%$\phi(x,t)$ and $\phi^{*}(x,t)$ , and therefore one can compute
%a new $\rho(x,t)=\left|\phi(x,t)\right|^{2}$ as well as $\phi_{f}(x)$
%and $\phi_{i}^{*}(x)$ . Although these equations conserve the total
%number of particles (space integral of $\rho(r,t)$), as a safety
%measure one must normalize $\rho(x,t)$ by requiring that its integral
%should be equal to that of $\rho_{i}$ . From eq.(\ref{eq:1d3}) and
%(\ref{eq:1d4}), one obtains a new $\phi_{i}(x)$ and $\phi_{f}^{*}(x)$,
%and the procedure can be iterated until convergence.

%We now look more specifically to the case of the ``echo'', namely
%the case when $\rho_{f}(x)=\rho_{i}(x)=\rho_{0}(x)$ . We want to
%compute the probability that the particles have the same final density
%profile at time $t_{f}$ as the initial one. Since the boundary condition
%on $\phi(r,0)$ is identical to that of $\phi^{*}(r,t_{f})$ we have
%\[
%\phi^{*}(x,t)=\phi(x,t_{f}-t)
%\]
%and thus
%
%\[
%\phi(x,0)\phi(x,t_{f})=\rho_{0}(x)
%\]
%and
%
%\[
%\rho(x,t)=\phi(x,t_{f}-t)\phi(x,t)
%\]
%We look specifically at the case when
%
%\[
%\rho_{0}(x)=\theta(x)
%\]
%

The corresponding probability is given by

\begin{eqnarray}
P(w_{f},t_{f}\left|w_{i},0\right.) & = & N!\exp\Big(-\frac{\beta g}{4}\int dx(\rho_{f}^{2}(x)-\rho_{i}^{2}(x))\Big)\label{proba1d}\nonumber \\
\\ & \times & \exp\Bigg\{-\frac{D\beta g}{2}\int_{0}^{t_{f}}dt\int dx\rho(x,t)\rho''(x,t)\nonumber \\
\\ & + & \frac{D\beta^{2}g^{2}}{2}\int_{0}^{t_{f}}dt\int dx\rho(x,t)\rho'{}^{2}(x,t)\nonumber \\
 & + & \int dx\left(w_{f}(x)\log\frac{w_{f}(x)}{\phi(x,t_{f})}+w_{i}(x)\log\frac{w_{i}(x)}{\phi^{*}(x,0)}\right)\Bigg\}
\end{eqnarray}

\begin{figure}[h]
\begin{subfigure} {.5\textwidth} 
\centering 
\includegraphics[width=1\linewidth]{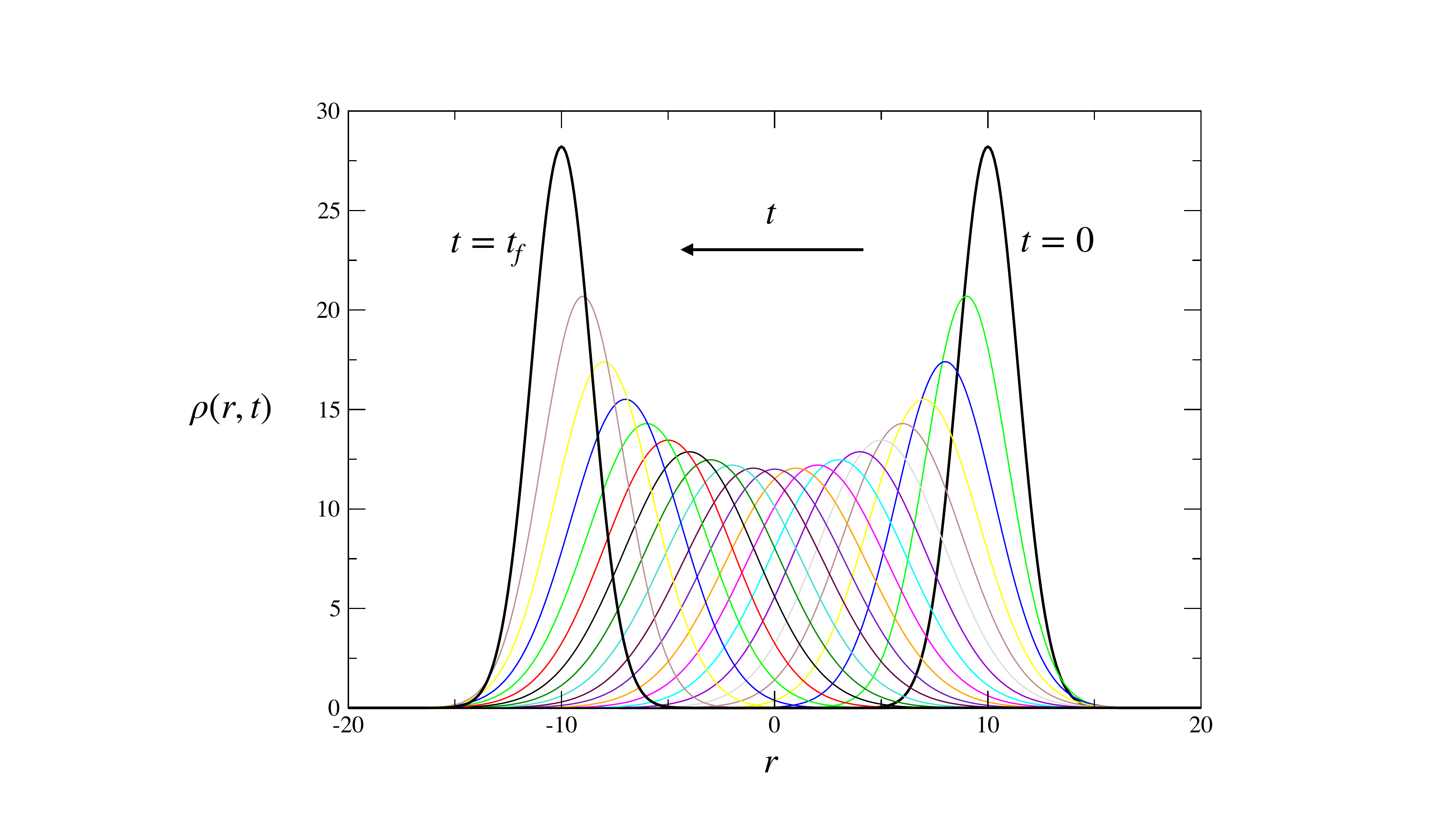}
\caption{$g=0$ Free case}
\end{subfigure}%
\begin{subfigure}{.5\textwidth} 
\centering 
\includegraphics[width=1\linewidth]{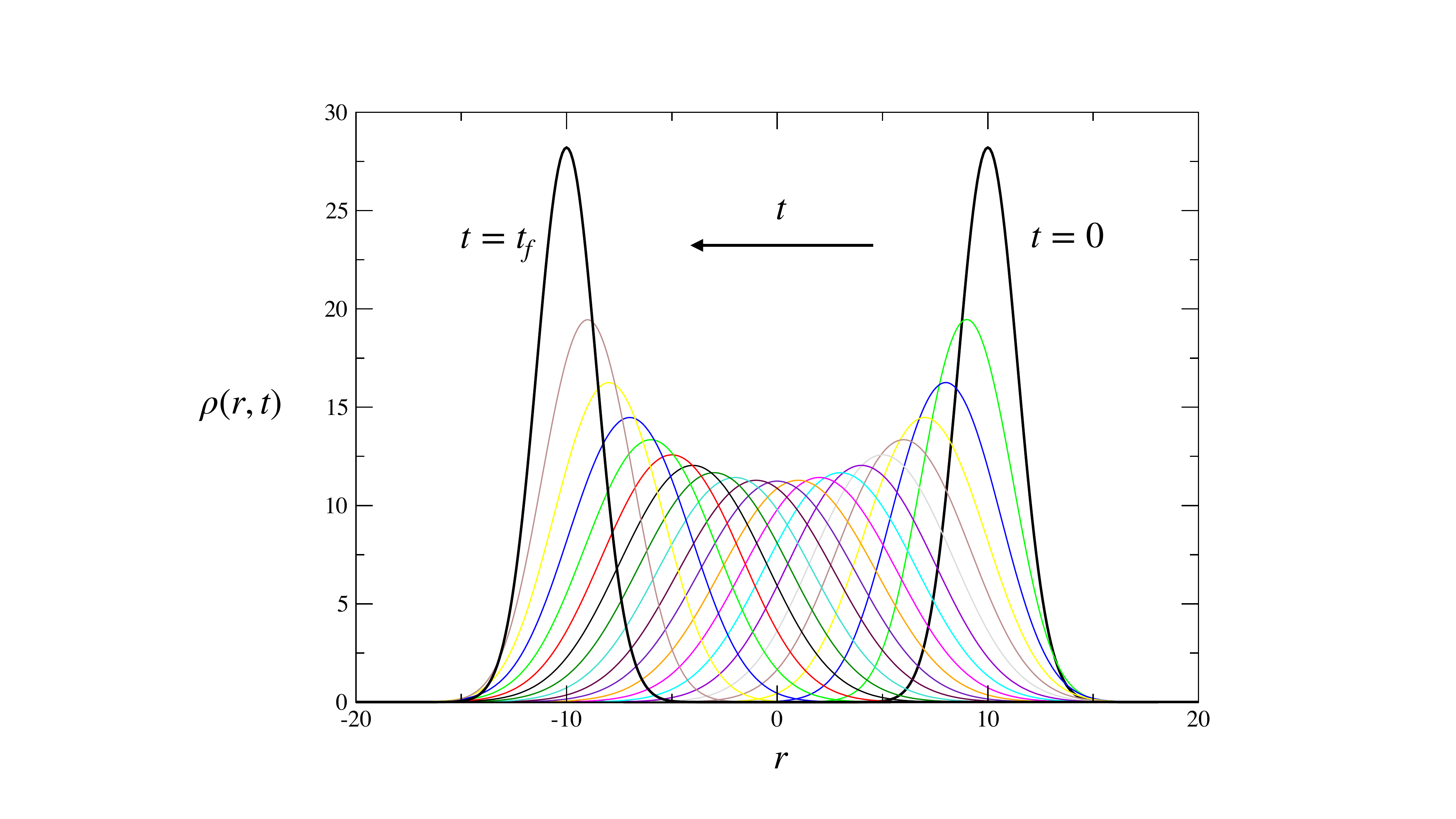}
\caption{$g=0.03$ Repulsive case}
\end{subfigure} 
\caption{Snapshot of the distribution of particles between the initial (right)
and final (left) distributions (thick black lines). The arrow denotes
the direction of increasing time}
\label{fig1} 
\end{figure}

To illustrate the method, we have solved numerically equations (\ref{eq:1d1})
and (\ref{eq:1d2}) for some typical values of $g$. We impose hard-wall
boundary conditions. The initial and final distributions are identical
Gaussians separated by a distance $d$ 
\begin{equation}
\rho_{i,f}(x)=\frac{N}{\sqrt{4\pi\sigma^{2}}}e^{-\frac{(x\mp d/2)^{2}}{2\sigma^{2}}}
\end{equation}
The number of particles is $N=100$, the distance $d=10$, diffusion
constant $D=T=2$. %\begin{figure}[h]
%\begin{
%	\centering
%	\includegraphics[width=0.5\linewidth]{./density0}
%	\includegraphics[width=0.5\linewidth]{./density0.03}	
%	\caption{
%Snapshot of the distribution of particles between the two boundary distributions for $g=0$ (left) and $g=0.03$ (right)}
%	\label{fig1}
%\end{figure}
We first study the case of no interaction $g=0$ (a) and repulsive $\delta$
interaction $g=0.03$ (b). The iteration method does not converge for larger values of $g$. Since the interaction energy of the system scales like $g \rho^2 V$ where $\rho$ is the average density of the system and  $V$ its volume, the natural scaling of $g$ is $g \sim 1/\rho^2$. In our numerical example, $N=100$ and the typical volume of the system is $V\sim 20$, which implies an average density $\rho\sim 5$. This means that $g$ should be of the order 0.04.
As can be seen in fig.(\ref{fig1}), the effect of repulsive interaction
is small, barely visible. This is expected at low densities. The repulsive effect of the interaction is masked by the entropic repulsion of the particles. 

\begin{figure}[h]
\centering 
\includegraphics[width=0.5\textwidth]{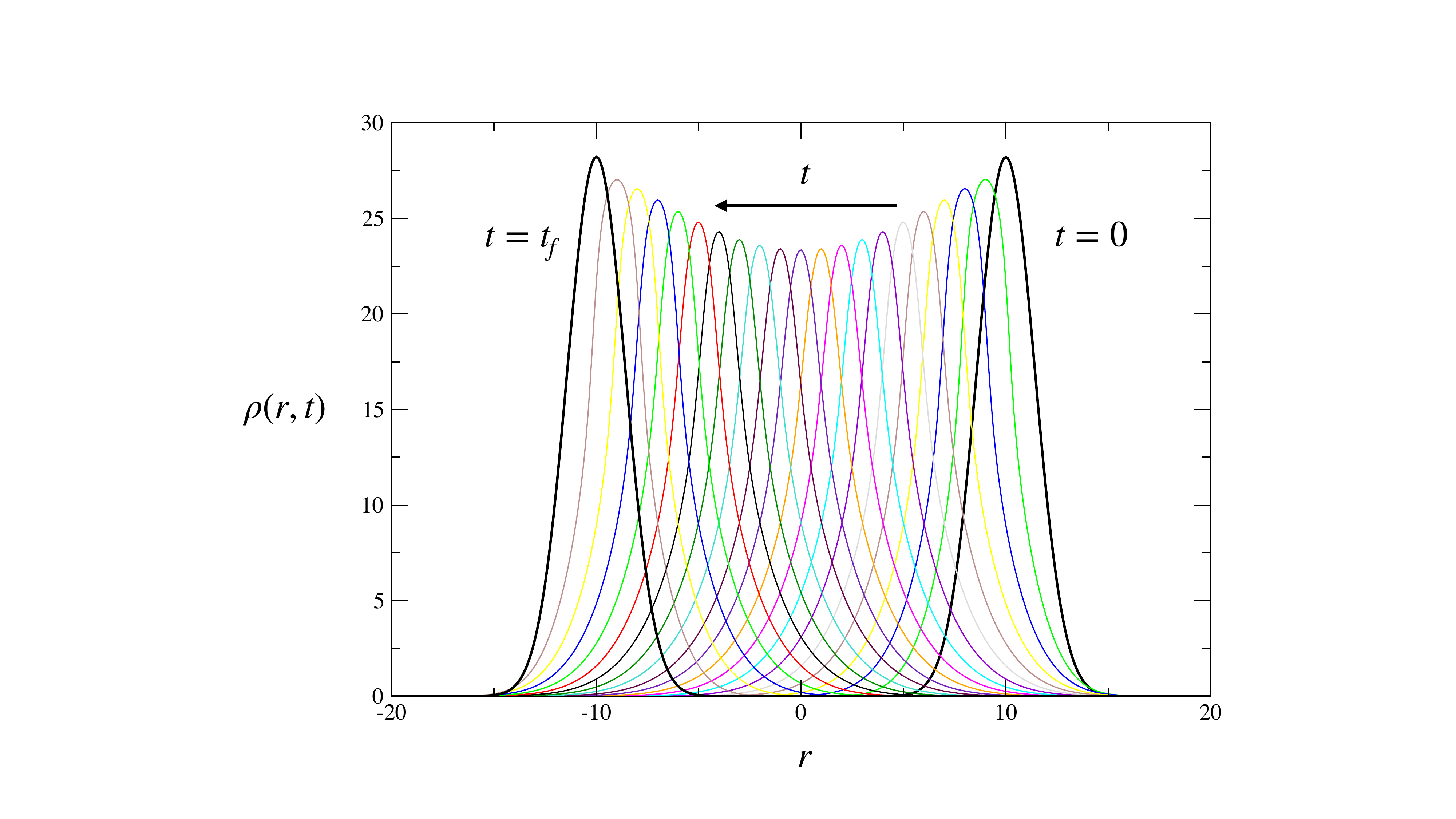}
\caption{ $g=-0.1$ Attractive case}
 \label{fig2} 
\end{figure}

The case of attractive interaction is more spectacular. In fig.(\ref{fig2}),
we plot the sequence of distributions between the two boundary distributions
for $g=-0.1$.   

As shown in fig.(\ref{fig2}), the distribution remains much more
concentrated than in the zero or repulsive case during the transition,
due to the attractive interactions. The data for the figures are available at \cite{fig}.
In all cases, the fixed-point
method does not converge for larger values of $g$.

In this article, we have generalized the concept of Schr\"odinger bridge
to the case of interacting particles. 
As there is no practical formulation of this problem as an optimal transport or control problem, it is useful to formulate it as a statistical field theory, which can be tackled with the usual tools of field theory.
We derived partial differential equations for the
most probable distribution path between two boundary distributions.
We have checked that these equations can be solved numerically by
a generalization of the Sinkhorn algorithm, at least in some simple
one-dimensional cases. This generalization should prove very useful
in the study of problems of non-equilibrium statistical physics of many-particle systems, such as the dynamics of phase transitions, the glass transition, protein folding, etc. However, the use of more realistic interactions to study such physical problems requires devising efficient numerical
algorithms to solve these equations. The author is currently working on searching for such algorithms.

Another ongoing project is to adapt this formalism to the real-time quantum Schr\"odinger
evolution, both for bosons and fermions. Here again, further studies are needed to devise reliable numerical methods for large coupling cases.

\begin{acknowledgements} The author wishes to thank Prof. K. Mallick
for illuminating discussions. 
\end{acknowledgements}


\begin{thebibliography}{10}
\bibitem{Schrod1} E.Schr\"odinger, \textit{Sitzungsberichte der Preussischen
Akademie der Wissenschaften} , Physikalisch-mathematische Klasse,
\textbf{144}, 144 (1931)

\bibitem{Schrod2} E.Schr\"odinger, \textit{Sur la théorie relativiste
de l'électron et l'interprétation de la mécanique quantique}, Annales
de l'I.H.P., tome 2, 4 (1932), 269-310.

\bibitem{Chetrite} R.Chetrite, P.Muratore-Ginanneschi and K.Schwieger,
Eur. Phys. J. H (2021) 46:28.

\bibitem{Leonard} C.Leonard, Discrete and Continuous Dynamical Systems,
\textbf{34},4, April 2014.

\bibitem{Foellmer} H. F\"ollmer,  (1986), Time reversal on Wiener space. In: Albeverio, S.A., Blanchard, P., Streit, L. (eds) Stochastic Processes. Lecture Notes in Mathematics, vol 1158. Springer, Berlin.

\bibitem{KL} S.Kullback and R.A.Leibler, \textit{On information and
sufficiency}, Annals of Mathematical Statistics. \textbf{22} (1):
79-86 (1951).

\bibitem{Aurell}E. Aurell, K. Gawedzki, C. Mejia-Monasterio, R. Mohayaee, P. Muratore-Ginanneschi, J. Stat. Phys. {\bf 147}, 487-505 (2012).

\bibitem{Saito} T. Van Vu and K. Saito, Phys. Rev X {\bf 13}, 011013 (2023).

\bibitem{SP}C.Leonard, Journal of Functional Analysis, \textbf{262}
4, 15 February 2012, 1879-1920.

\bibitem{TP}H.Orland, J. Chem. Phys. \textbf{134}, 174114 (2011).

\bibitem{OT}R.Otto and C.Villani, Journal of Functional Analysis,
\textbf{173}, 2, 1 June 2000, 361-400.

\bibitem{ML}A.Genevay, G.Peyré, and M. Cuturi, \textit{Learning generative
models with Sinkhorn divergences} in International Conference on Artificial
Intelligence and Statistics,1608-1617, PMLR, 2018.

\bibitem{CV}G.Peyr\'e and M.Cuturi, \textit{Computational Optimal Transport:
With Applications to Data Science}, Foundations and Trends in Machine
Learning: \textbf{11}:5-6, pp 355-607 (2019).

\bibitem{finan} M.Beiglb\"ock, P.Henry-Labordère and F.Penkner, Finance
Stoch \textbf{17}, 477-501 (2013).

\bibitem{LD} N.G.Van Kampen, \textit{Stochastic Processes in Physics
and Chemistry}. North-Holland (2007).

\bibitem{HO} R.Elber, D.E.Makarov and H.Orland, \textit{Molecular
Kinetics in Condensed Phases}, Wiley (2020).

\bibitem{Zwan} R.Zwanzig, \textit{Nonequilibrium Statistical Mechanic},
Oxford University Press (2001).

\bibitem{FH} R.P.Feynman and A.R.Hibbs, \textit{Quantum Mechanics
and Path Integrals} McGraw-Hill (1965).

\bibitem{QM} C.Cohen-Tannoudji, B.Diu and F.Laloe, \textit{Quantum
Mechanics} Wiley (1977).

\bibitem{Neg_HO} J.W.Negele and H.Orland, \textit{Quantum Many-Particle
Systems} (Advanced Books Classics), Perseus Books (1998).

\bibitem{Peskin} M. Peskin and D. Schroeder, \textit{An Introduction to Quantum Field Theory}, Westview Press (1995).

\bibitem{Sink} M.Cuturi in \textit{Advances in Neural Information
Processing Systems}, C.J. Burges, L. Bottou, M. Welling, Z. Ghahramani
and K.Q. Weinberger ed., \textbf{26} (2013)

\bibitem{Conv} P.A.Knight, \textit{The Sinkhorn-Knopp Algorithm:
Convergence and Applications}, SIAM Journal on Matrix Analysis and
Applications, \textbf{30} (1), 261-275 (2008).

\bibitem{fig} 
\url{https://osf.io/ybrkw/?view\textunderscore only=fb7005bf93254bcab7de13e94dfae08a}
%{https://osf.io/ybrkw/?view\textunderscore only=fb7005bf93254bcab7de13e94dfae08a}

\end{thebibliography}
\end{document}